\documentclass[mathleft
]{an}
\usepackage{graphicx}
\usepackage{times}
\begin{document}

\Pagespan{789}{}
\Yearpublication{2006}%
\Yearsubmission{2005}%
\Month{11}%
\Volume{999}%
\Issue{88}%

\title{Probing the temporal and spatial variations of dust emission in the protoplanetary disk of DG\,Tau with VLTI/MIDI -- Preliminary results}

\author{K.\'E. Gab\'anyi\inst{1}\fnmsep\thanks{Corresponding author:
  \email{gabanyi@konkoly.hu}\newline}
\and L. Mosoni\inst{1} \and A. Juh\'asz\inst{2} \and P. \'Abrah\'am\inst{1} \and Th. Ratzka\inst{3} \and N. Sipos\inst{4,5} \and R. van Boekel\inst{6} \and W. Jaffe\inst{4}
}
\titlerunning{DG\,Tau with VLTI/MIDI: Preliminary results}
\authorrunning{K. \'E. Gab\'anyi, L. Mosoni, et al.}
\institute{
MTA Research Centre for Astronomy and Earth Sciences, Konkoly Thege Mikl\'os Astronomical Institute, H-1121 Konkoly Thege Mikl\'os \'ut 15-17, Budapest, Hungary
\and 
Leiden Observatory, Leiden University, PO Box 9513, 2300 RA Leiden, The Netherlands
\and
Universit\"ats-Sternwarte M\"unchen, Ludwig-Maximilians-Universit\"at M\"unchen, Scheinerstr 1, 81679 M\"unchen, Germany
\and
Institute for Astronomy, ETH Z\"urich, Wolfgang-Pauli-Strasse 27, 8093 Z\"urich, Switzerland
\and
Universit\"at G\"ottingen, Institut f\"ur Astrophysik, Friedrich-Hund-Platz 1, D-37077 G\"ottingen, Germany
\and 
Max-Planck-Institut f\"ur Astronomie, Heidelberg 69117, K\"onigstuhl 17, Germany}
\received{Nov 2012}
\accepted{ 2012}
\publonline{later}


\abstract{
The signatures of dust processing and grain growth - related to the formation of rocky planets - are easily seen in mid-infrared
spectral features. One important diagnostic tool in this context is the silicate feature in the spectra of young stellar objects (YSO). The low-mass YSO,
DG\,Tau shows unique temporal variations in its silicate feature. We conducted multi-epoch observations of DG\,Tau with the
MID-Infrared Interferometric instrument of the Very Large Telescope Interferometer to obtain the spectra of the inner and outer
disk regions in order to investigate where the previously reported variations of the silicate feature originate from. Here we present
the preliminary results of the first two epochs of observations. We found that on a time-scale of two months, the source showed significant brightening. At the same time the mid-infrared emitting region expanded. While the identification of the silicate feature is difficult, our results qualitatively agree with the scenario explaining the varying silicate feature with dust lifted up above the disk.}

\maketitle

\section{Introduction}

Silicate grains in the protoplanetary accretion disks around young stellar objects (YSOs) play a key role in the
evolution of circumstellar disks and eventually in the formation of planetary systems. The signatures of dust processing
can be seen in the mid-infrared (MIR), where the shape and strength of dust features are determined by the composition and the
physical properties of the radiating dust grains. High spatial resolution MIR spectroscopy of Herbig Ae stars
(van Boekel et al. 2004) with the MID-Infrared Interferometric instrument (MIDI, Leinert et al. 2003) of the Very Large
Telescope Interferometer (VLTI) showed for the first time that the spatial distribution of dust species and their properties
are not homogeneous in the circumstellar disks. These observations revealed that the inner disk regions contained more
crystalline silicates than the outer regions, and in two cases out of three, the larger grains were concentrated in the
inner disks. 

DG\,Tau is a low-mass YSO (mass $\sim 0.7M_\odot$, spectral type K3, White \& Hillenbrand 2004) known to possess a thick
and actively accreting disk. Variations in its silicate emission feature were observed by e.g., Bary, Leisenring \&
Skrutskie (2009) (also see references therein). Bary et al. (2009) analysed {\it Spitzer} Infrared Spectrograph (IRS)
spectra taken on 9 epochs between 2004 March and 2007 March. They reported significant variations observed on
time-scales from a week to months. The short timescale of the variations imply that the dust obscuration occurs on
dynamical timescales consistent with the motion of the dust within a disk, at radii of $\le1$AU. Thus, obscuration by
a clumpy dust envelope can be ruled out. The authors propose further scenarios to explain the time variability of the
silicate feature. For example, following Muzerolle et al. (2009), the variations both on short and long timescales can be explained
by a disk shadowing effect, where the puffed-up inner rim of the dust disk casts a shadow onto the disk surface. We note, that since DG~Tau is seen almost face-on (inclination $\le 30^\circ$ - see Isella, Carpenter \& Sargent (2010) and references therein), the outer disk likely cannot cover the inner regions and so cause the absorption.
Alternatively, turbulent mixing in the disk (Turner, Carballido \& Sano 2010, Balsara et al. 2009) or 
dusty disk winds (Tambovtseva \& Grinin 2008, Vinkovi\'c \& Jurki\'c 2007) can be responsible for lifting up
material above the disk, hence producing self-absorption of the emitting region. 

Most of the published MIR spectra of DG~Tau show the silicate in emission. However, there is one case where the feature
is in absorption (Sitko et al. 2008). Since the disk shadowing effect cannot produce a silicate feature in absorption, 
this suggests that dredged-up cool dust can be responsible for the silicate absorption.
However, a sufficiently large amount of dust must be lifted up from the disk surface and cooled down to $\approx 150$\,K. Otherwise it will produce additional silicate emission. In this latter case, the dust above the disk can increase the silicate emission which can later decrease as the dust settles into the disk. Therefore this scenario, depending on the amount of dust above the disk, can alone explain all kinds of variations of the silicate feature.

We conducted multi-epoch observations of DG\,Tau with the VLTI/MIDI. We aim to study the different dust species at different spatial scales in the disk. Furthermore, by utilizing the variability of the source, we intend to determine the location of the variation of the
silicate emission. Additionally, with the help of MIDI observations we can
constrain the mechanism behind the variation of the silicate feature. In this paper, we report on the preliminary
results of the first two epochs of observations.

\section{Observations and data reduction}

MIDI combines the signal from two 8.2\,m Unit Telescopes (UTs) or from two 1.8\,m Auxiliary Telescopes (ATs), and
provides interferometric data in the wavelength range of $8-13\,\mu$m. Three epochs of observations (with UTs) of DG\,Tau
have been already performed with VLTI/MIDI in HIGH-SENS mode (for details of this observing mode see Chesneau 2007), and
additional epochs are scheduled for the end of 2012 with both UTs and ATs. Here, we only present the first two
epochs conducted on 2011 October 10 and on 2011 December 13 using two different baselines. (Further details of the
observational setups are given in Table~\ref{tab:obs}.) The data reduction of the third epoch has just started. The narrow
range of position angles ($30^\circ$ to $45^\circ$) allows us to study the structure and dust distribution along a
well defined direction in the disk. 

\begin{table}
\caption{Details of the VLTI/MIDI observations of DG\,Tau. Position angle is measured from North through East.}
\label{tab:obs}
\begin{tabular}{cccc}\hline
Epoch & Baseline & Projected & Position angle\\ 
 & & baseline length & (deg) \\
\hline
2011-10-10 & UT1-UT2 & 33\,m & 29\\
2011-10-10 & UT1-UT3 & 75\,m & 45\\
2011-12-13 & UT1-UT2 & 37\,m & 34\\
2011-12-13 & UT1-UT3 & 56\,m & 35\\
\hline
\end{tabular}
\end{table}

The obtained data sets consist of 
the N-band ($8-13\,\mu$m) low resolution spectra ($\lambda/\Delta\lambda \approx 30$), and the interferometric 
(correlated) spectra of the target in the same wavelength range with the same spectral resolution. 
The correlated spectra are heavily dominated by the inner region of the circumstellar disk, thus it can be considered as
the spectra of the inner region of the target, i.e. the inner disk of DG~Tau. 
The typical size of this region can be estimated by the angular resolution of the observations. 
With the shortest baseline we reach an angular resolution of $\approx 60$\,mas, corresponding to 8\,AU at
the distance of Taurus ($140$\,pc). 

In the data reduction we followed the general processing scheme as described in, e.g. Leinert et al. (2004) and Ratzka
et al. (2007). MIDI data can be reduced in two independent ways: with the MIDI Interactive Analysis (MIA) package, which
uses the power spectrum method, and the Expert Work Station (EWS) package, which is based on a coherent, linear
averaging method (Chesneau 2007)\footnote{The EWS+MIA package can be obtained from:
\texttt{www.strw.leidenuniv.nl/$\sim$jaffe/ews/index.html}}. That way we could cross-check the results of the two data
reduction processes. We found that within the errors the output of the two data reduction processes agreed.

The calibrator used in all observing runs was HD\,27482. We used the spectra available in the MIA+EWS program package
(version 1.7) within the \texttt{vboekelbase()} database to calculate its flux density at $10\,\mu$m, which is $6.45$\,Jy. From
the same database, we obtained the source size, $2.27$\,milli-arcsecond (mas), thus it is an unresolved point source 
at the resolution provided by MIDI. 

The total spectra of DG\,Tau at the two epochs are displayed in Fig.~\ref{spectra}. The atmospheric ozone absorption
feature centered around $9.7\,\mu$m could not be perfectly calibrated, so we decided to disregard points in the
affected wavelength regime, between $9.4\,\mu$m and $10.0\,\mu$m in both the total and the correlated spectra. The errors of the total and correlated spectra are wavelength dependent. The uncertainty in the total spectra is $\sim 5$\,\% (i.e., $\pm 0.3$\,Jy and $\pm 0.2$\,Jy in the first and second epoch, respectively) at the shortest wavelengths (i.e., at $\sim 8.5\,\mu$m). The errors increase with wavelength and reach values as high as $25$\,\% (i.e., $\pm 1$\,Jy and $\pm 1.4$\,Jy in the first and second epoch, respectively) at the longest wavelength (at $12.5\,\mu$m). 
The errors of the correlated spectra on the shorter baseline (UT1-UT2, see Fig.~\ref{corrflux12}) are between $7\,\%$ and $14\,\%$ (i.e., $\pm 0.2$\,Jy and $\pm 0.4$\,Jy) in the first epoch and between $3\,\%$ and $12\,\%$ (i.e., $\pm 0.1$\,Jy and $\pm 0.3$\,Jy) in the second epoch. The errors on the longer baseline (UT1-UT3, see Fig.~\ref{corrflux13}) are significantly higher in both epochs, between $10\,\%$ and $27\,\%$ (i.e., $\pm 0.3$\,Jy and $\pm 0.8$\,Jy). Note that in principle, the errors of the correlated spectra (if derived by the method employed by EWS, as in this case) may have lower values than those of the total spectra thanks to the fact that the contribution from the uncorrelated sky background cancels out (for details see Ratzka et al. 2007).

\section{Results}

\begin{figure}
\includegraphics[bb=80 40 560 715,angle=270,width=80mm,clip]{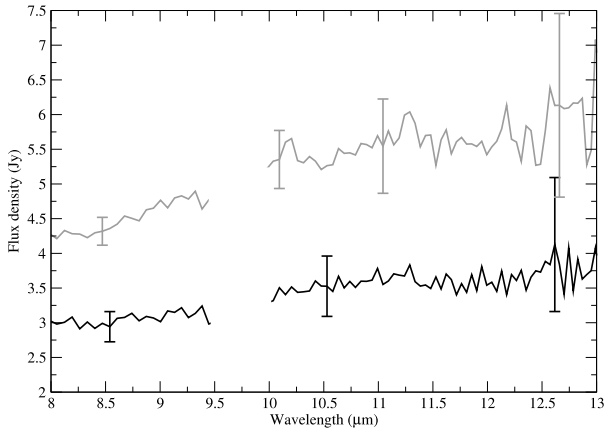}
\caption{The total spectra of DG\,Tau in epoch 1 (black curve) and in epoch 2 (grey curve). For clarity, we show only 
representative error bars in each spectrum.}
\label{spectra}
\end{figure}

The source showed significant brightening in its total spectrum from the first to the second epoch (Fig.~\ref{spectra}). We expect that the
MIR emitting region will be more extended when the central source is brighter, where we assume that the most of the 
MIR excess originates from the re-radiation of the central illumination by the dusty disk. The source of the 
overall brightening might be the stellar surface or the inner gas disk and is likely related to a varying mass accretion
rate.

\begin{figure}
\includegraphics[bb=80 40 560 715,angle=270,width=80mm,clip]{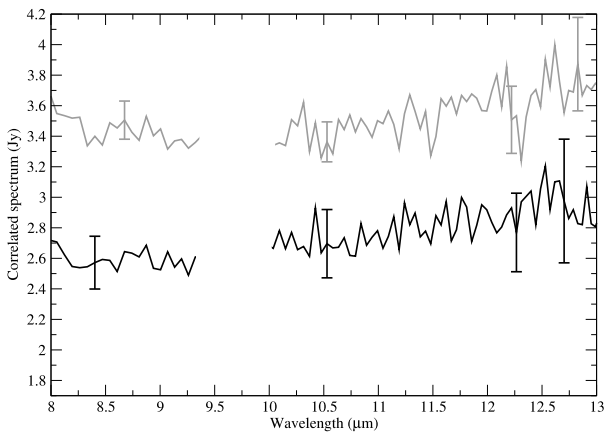}
\caption{The correlated spectra of DG\,Tau on the baseline UT1-UT2 in epoch 1 (black curve) and in epoch 2 (grey curve).}
\label{corrflux12}
\end{figure}

\begin{figure}
\includegraphics[bb=80 40 560 715,angle=270,width=80mm,clip]{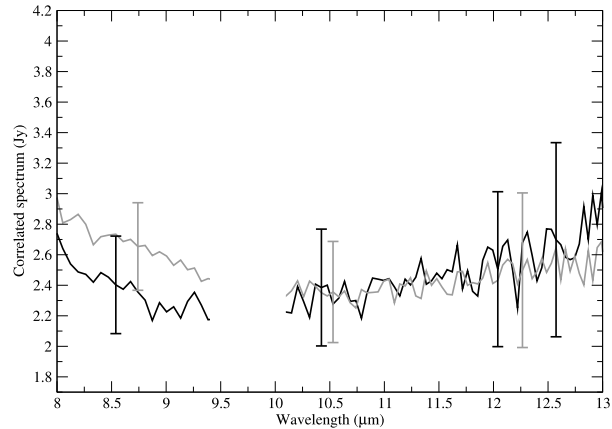}
\caption{The correlated spectra of DG\,Tau on the baseline UT1-UT3 in epoch 1 (black curve) and in epoch 2 (grey curve).}
\label{corrflux13}
\end{figure}

In Figs.~\ref{corrflux12} and ~\ref{corrflux13}, the correlated spectra are shown for the shorter UT1-UT2 baseline and the
longer UT1-UT3 baseline, respectively. For the shorter baseline, a clear brightening of the inner disk can be seen from 2011
October to December, similarly to the total spectrum. (The similarity of the baseline vectors allows the direct comparison.) However, no
significant variations can be seen at the longer baseline, although we note that the measurement errors are much higher in these
observations. 

\begin{figure}
\includegraphics[bb=80 40 560 715,angle=270,width=80mm,clip]{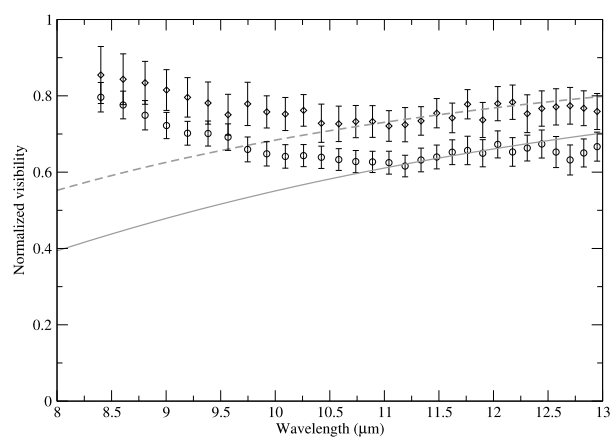}
\caption{Normalized visibilities for DG\,Tau measured with UT1-UT2 baseline. 
Diamonds represent observation done in 2011 October, circles represent observation done in 2011 December. Grey curves show the Gaussian angular size estimates (dashed - 17\,mas, dotted - 19\,mas FWHM).}
\label{vis}
\end{figure}

The ratio of the correlated and total spectra gives the spectrally resolved visibilities. 
For an unresolved object the visibilities are 1, lower values indicate a resolved structure. 
The normalized visibilities are displayed for the UT1-UT2 baseline observations in Fig.~\ref{vis}. The source is
clearly resolved. To estimate the size of the MIR emitting region, we approximated the
observed visibilities at the longer wavelengths (above $10.5\,\mu$m) with the model visibilities derived from a simple Gaussian
brightness distribution. This way we can estimate the size of the inner region responsible for the observed correlated
spectra. The resulting full-width half maximum (FWHM) sizes of the Gaussian models for the different baselines range
between 12 and 19\,mas.
This simple model shows that the MIR emission, longward of $10.5\,\mu$m of the correlated spectra clearly originates from the
inner $\sim 2$\,AU region of the disk. Interestingly, in the second epoch the visibilities longward $9\,\mu$m are significantly below the
visibility values measured at the first epoch. This can be understood by considering the increase in the total
MIR brightness of the source shown by the total spectra. As the source brightened, the MIR emitting region expanded,
which in turn can be observed as a decrease in the visibilities. 

The simple Gaussian model cannot describe adequately the observed visibilities at short wavelengths. At these wavelengths (shorter than $\sim 10.5\,\mu$m), the observations also sample a large contibution from a much hotter region, the inner rim of the disk even closer to the star. Thus the effective size of the disk appears much smaller at these wavelengths than expected from a one-component geometrical Gaussian model (Leinert et al. 2004).

While there is a slight sign of silicate in emission in the total spectrum in the second epoch, in the correlated
spectrum there is no such indication. On the contrary, the correlated spectra may indicate a silicate feature in
absorption both at the long and short baselines. (As a consequence, the N-band spectrum of the outer disk should show 
the silicate feature in emission.) Interestingly, the absorption is deeper in the correlated spectra of the
longer baseline, thus in the spectra of the innermost region of the disk. 
The previously reported variations in the silicate feature can be explained via dredged-up material (either by disk wind or turbulent mixing) above the disk. The presence of cool dust above the disk is the easiest explanation for the absorption seen in the correlated spectra of our MIDI observation.
However, to confirm the shape of the spectra and the silicate feature, one needs to achieve a better signal-to-noise ratio and conduct a careful continuum subtraction. Fitting the continuum is particularly difficult. Firstly, because the silicate feature is broad compared to
the wavelength range of MIDI, and  secondly, because of the large errors inherent of ground-based MIR
observations.

\section{Summary and outlook}

We observed the low-mass YSO DG\,Tau with the VLTI/MIDI instrument to investigate the spatial distribution of the dust
grains in its protoplanetary disk, and to constrain the location and  mechanism behind the temporal variation seen in
the silicate feature. Here, we showed the preliminary results of the first two epochs of our observational campaign. 
We observed variations both in the total and correlated flux density of the source on a timescale of two months. The measured visibilities showed that the source was more expanded when brighter; this can be explained by the expansion of the MIR emitting region.
Different scenarios have been proposed in the literature to explain the variations of the silicate feature. Our data supports the scenario, in which the dust lifted from the disk surface is responsible for the varying silicate emission. Additional data obtained in two more epochs hopefully will help to understand in detail the physical processes in the close environment of DG~Tau.

The next steps are (i) to finalize the data reduction and achieve the required data quality; (ii) work
out the method of continuum subtraction and (iii) carry out the study of the silicate feature. This latter can either be
done by comparing the continuum-subtracted spectra to templates (see e.g., van Boekel et al. 2004) or by fitting the dust
composition (see Juh\'asz et al. 2009).

\acknowledgements
Based on observations made with ESO telescopes at the Paranal Observatory under program ID 088.C-1007. The authors thank the ESO/VLTI staff for executing the observations in service mode. The research leading to these results has received funding from the European Community's Seventh Framework Programme under Grant Agreement 226604. The authors are thankful for the support of the Fizeau Exchange Visitor Program through the European Interferometry Initiative (EII) and OPTICON (an EU funded framework program, contract number RII3-CT-2004-001566). Financial support from the Hungarian OTKA grants K101393 and NN102014 are acknowledged.


\end{document}